\documentclass[a4paper]{spie}  

\usepackage{amsmath,amsfonts,amssymb,aas_macros}
\usepackage{graphicx}
\usepackage{url}

\usepackage[colorlinks=true, allcolors=blue]{hyperref}

\title{FlyEye Ground-Based Telescope: 
Unveiling New Frontiers in Astronomical Science
}

\author[a,b]{Carmelo Arcidiacono}
\author[a]{Matteo Simioni}
\author[a,b,d]{Roberto Ragazzoni}
\author[c]{Piero Gregori}
\author[c]{Paolo Lorenzi}
\author[c]{Francesco Cerutti}
\author[c]{Roberto Ziano}
\author[c]{Matteo Bisiani}
\author[c]{Roberta Pellegrini}
\author[c]{Andrea Guazzora}
\author[e]{Silvano Pieri}
\author[a,b]{Marco Dima}
\author[a,b,d]{Silvio Di Rosa}
\author[a]{Simone Zaggia}
\author[a.b]{Jacopo Farinato}
\author[a,b]{Demetrio Magrin}
\author[a]{Andrea Grazian}
\author[a]{Marco Gullieuszik}

\affil[a]{INAF Osservatorio Astronomico di Padova, Vicolo dell'Osservatorio 5, I-35122, Padova, Italy}
\affil[b]{ADONI, Laboratorio Nazionale di Ottica Adattiva}
\affil[c]{OHB Italia S.p.A., Via Gallarate 150, I-20151, Milano, Italy }
\affil[d]{Dipartimento di Fisica e Astronomia, Universit\'a degli Studi di Padova, Vicolo dell’Osservatorio 3, I-35122 Padova, Italy}
\affil[e]{Studio S.O.M.E. di Pieri Stefano SAS}
\authorinfo{Further author information: (Send correspondence to C.A.)\\C.A.: E-mail: carmelo.arcidiacono@inaf.it, Telephone: +39 049 829 3414}

\begin{document} 
\maketitle

\begin{abstract}
The FlyEye design makes its debut in the ESA’s NEOSTEL developed by OHB-Italia. This pioneering FlyEye telescope integrates a monolithic 1-meter class primary mirror feeding 16 CCD cameras for discovering Near-Earth Object (NEO) and any class of transient phenomena. OHB-Italia is the prime contractor, receiving extended support from the Italian National Institute for Astrophysics (INAF) in the ESA’s NEOSTED program's integration and testing. The FlyEye distinctive design splits the Field of View into 16 channels, creating a unique multi-telescope system with a panoramic 44 square degree Field of View and a seeing-size pixel-scale, enabling NEOs detection down to apparent magnitudes 21.5 insisting on a 1m diameter spherical mirror. 
The scientific products of a similar FlyEye telescope can complement facilities such as Vera Rubin (former LSST) and ZTF. The FlyEye has the ability to survey two-thirds of the visible sky about three times per night can revolutionize time-domain astronomy, enabling comprehensive studies of transient phenomena, placing FlyEye in a new era of exploration of the dynamic universe. Efforts to develop automated calibration and testing procedures are keys to realizing this transformative potential.	 
\end{abstract}


\section{INTRODUCTION}
\label{sec:intro}  
\begin{figure}
  \centering
  \includegraphics[width=0.9\linewidth]{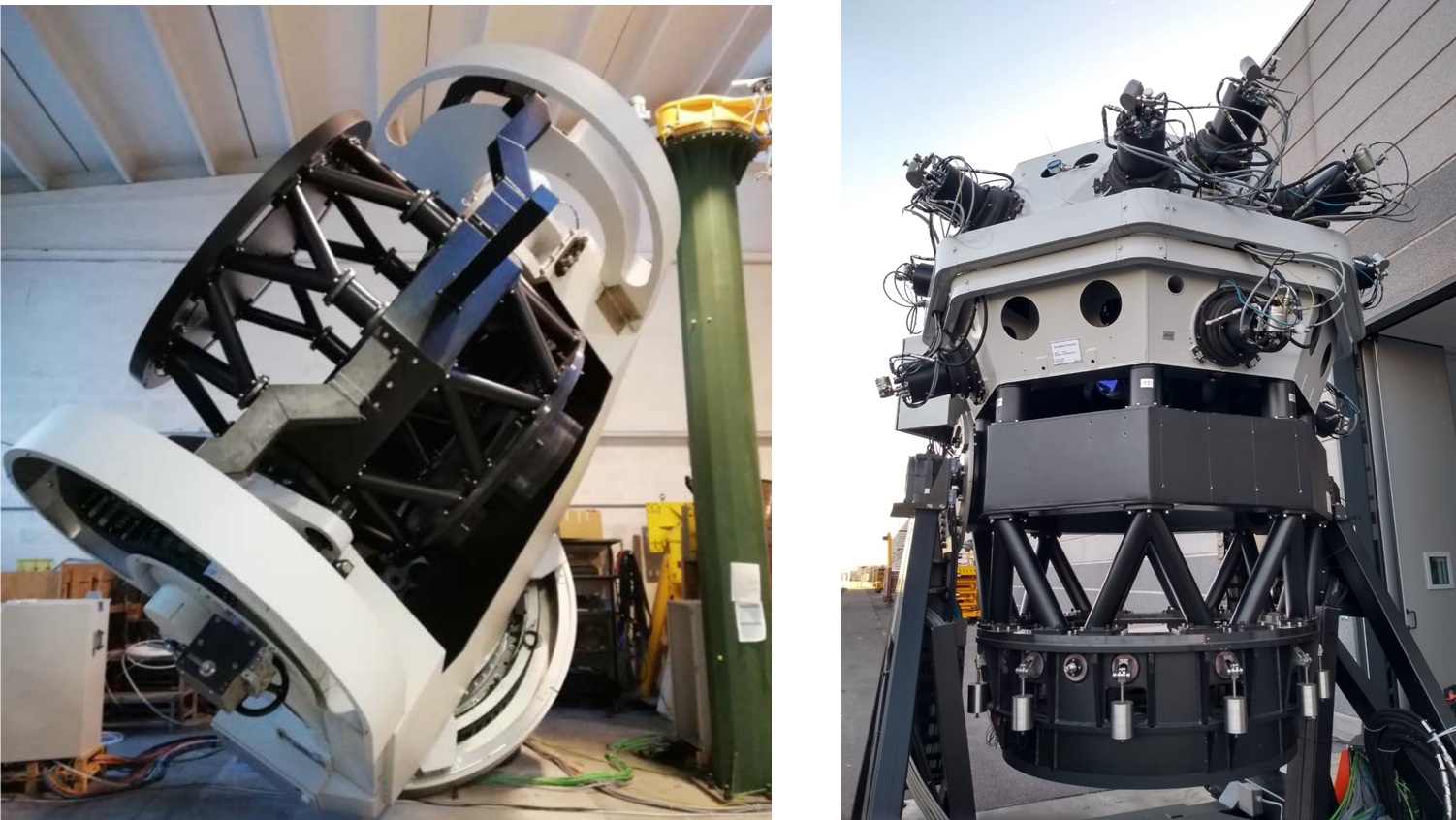}
      \caption{ESA’s NEOSTEL first FlyEye. The equatorial mount in Verona and the Telescope optics mounted on elevation only mount in Turate.}
         \label{fig:telescope}
  \end{figure}

The first phase of the NEOSTEL\cite{cibinNEOSTELTelescopeDetail2016} program has been nearly completed with the telescope accepted in-factory. Throughout 2021, the FlyEye\cite{gregoriFlyeyeTelescopeDesign2023} Telescope saw improvements in its optomechanical design and performance optimisation, including the successful upgrade of the FlyEye Equatorial Mount, Figure~\ref{fig:telescope}.
In 2023, the FlyEye cameras were upgraded by installing getter pumps to improve their vacuum performance\cite{gregoriFlyeyeTelescopeStudy2023a}, which is crucial for maintaining the cameras' operational efficiency. An intermediate commissioning and science verification campaign is expected to take place within 2024. This will integrate the optics with the Mount at the ASI Space Geodesy Centre located in Matera\cite{diceccoCommissioningScienceVerification2021,diceccoObservationalSitesCharacterization2018}, Italy, ahead of the final installation at the ESA-designated observation site in Monte Mufara, Sicily, which is under development directly by ESA, with EIE Group as the prime contractor\cite{marchioriNEOSTEDInfrastructuresFastest2020}.

From a scientific perspective, the on-sky performance of the current NEOSTEL - FlyEye telescope will bring detailed info for the development of a dedicated FlyEye for astronomical science.
Following the integration and initial testing in Matera, the FlyEye Telescope is scheduled for final installation at the observation site currently being developed by ESA. This installation aims to enhance its capabilities for near-earth object detection and tracking, contributing to the ESA Space Safety Programme. Its primary objective is to detect Near Earth Objects (NEOs)\cite{perozziEfficientDeploymentStrategy2021} that are on a collision course with Earth, providing crucial advance notice of potential impacts within weeks or days.
\section{Optical Design of the FlyEye Telescope}

The FlyEye telescope features a unique optical design optimized for wide-field astronomical surveys, particularly suited for near-Earth object (NEO) detection. The optical system is engineered to provide a wide field of view (FoV) while maintaining high image quality throughout the field, which is critical for reliable detection and tracking of objects in space.

\subsection{Optical Configuration}

The FlyEye telescope utilizes a multi-lens design that divides the incoming light into 16 separate channels\cite{arcidiaconoDevelopmentFirstFlyeye2020a}, each equipped with its own CCD camera. This configuration allows for simultaneous observations over a large portion of the sky, approximately 44 square degrees, making it highly effective for rapid sky surveys.

\begin{itemize}
    \item \textbf{Primary Mirror:} The telescope is equipped with a 1-meter class primary mirror, which is monolithic and provides a stable and uniform reflective surface to capture incoming light efficiently.
    \item \textbf{Lens System:} Each of the 16 channels features a refractive lens system designed to minimize optical aberrations such as chromatic aberration and astigmatism, ensuring sharp images across the CCD sensors.
    \item \textbf{CCD Cameras:} The CCD cameras are high-sensitivity devices with fine pixel scales to detect even the faintest objects in the sky. The cameras are cooled to reduce noise, enhancing the quality of the captured images.
\end{itemize}

This optical design not only enables the FlyEye telescope to perform comprehensive sky surveys but also ensures that it can quickly adapt to different observing conditions, maintaining a high level of operational flexibility and efficiency.

ESA, in collaboration with the Italian Space Agency (ASI) identified the location for the installation of the first NEOSTEL prototype on top of the Mount Mufara on the Madonie Chain in Sicily\cite{buggeyDevelopingActiveTrap2022}.

\subsection{ASTROCAD Camera Characteristics}
\label{sec:astrocad}
On the NEOSTEL, each of the 16 channels is equipped with an ASTROCAD camera\cite{gregoriFlyeyeTelescopeDesign2023}, based on astronomical grade CCD from e2v Teledyne. The fast two-meter effective focal length, \( F\# = 2 \), matches on the \( 15 \, \mu\text{m} \) pixels, a \( 1.5 \, \text{arcseconds} \) pixel scale. The camera works in the \( 0.47 \, \mu\text{m} \) to \( 0.78 \, \mu\text{m} \) spectral range. On a dedicated FlyEye for astronomical science, the detector model and final effective focal ratio will be possibly tuned to match scientific requirements. See table~\ref{tab:ccd_specs}.

\begin{table}[h]
\centering
\caption{Characteristics of the CCD System}
\label{tab:ccd_specs}
\begin{tabular}{|l|l|}
\hline
\textbf{Characteristic} & \textbf{Detail} \\ \hline
Sensor Type & Astronomical grade CCD (e2v Teledyne, model CCD231-84-BI) \\ \hline
Cooling System & Operated at -45°C using a Thermoelectric Cooler (TEC) \\ \hline
Vacuum Chamber & Sealed chamber to prevent frost and minimize heat exchange, \\ & enhanced with getter pumps installed in 2023 \\ \hline
Field of View & 2.9 square degree each, 44 square degree total \\ \hline
Pixel Scale & 1.5 arcsec/px \\ \hline
Quantum Efficiency & $>$95\% peak, $>$75\% mean over a 470-780nm spectral window \\ \hline
Readout noise & 2e RMS maximum for a readout frequency of 50kHz \\ \hline
\end{tabular}
\end{table}

\section{OBSERVING STRATEGIES}
\label{sec:OBSERVING-STRATEGIES}
This comprehensive coverage is achieved through a unique optical system that divides the image into 16 subimages. FlyEye's design features a 1-meter class primary mirror and 16 CCD cameras, each mounted on a Secondary Optics Tube. The combined field of view is 44 square degrees, allowing it to capture the image of the visible night sky on a night. Furthermore, the system is designed with a fill factor 100\% to ensure that there are no instrumental gaps in the focal plane, at limited resolutions (1.5arcsec/px). I.e. The main mission of the NEOSTEL telescope is to survey the sky for unidentified Near-Earth Objects (NEOs): With exposure times set at 40 seconds, the telescope has the capacity to identify objects up to a limiting magnitude of 21.5. At this cadence, the FlyEye(s) can survey the visible sky in one night.

\section{Complementing Existing Facilities}
A dedicated FlyEye telescope can strategically work in synergy with other major astronomical projects, notably the Vera Rubin Telescope\cite{collaborationScientificImpactVera2020} (9.5 square degrees) and the Zwicky Transient Facility\cite{reileyOpticalDesignZwicky2017,dekanyInitialPerformanceZwicky2018} (ZTF, 47 square degrees). This integration can extend capabilities of this existing and planned observational infrastructures, by adding additional cadence to the visible sky survey or additional Earth longitude coverage.
The Vera Rubin Observatory Legacy Survey
of Space and Time is set to conduct its operation, which will catalogue billions of objects and is expected to produce a vast amount of data aimed at understanding dark matter and dark energy. Although the Vera Rubin telescope provides broad coverage, its survey cadence may miss transient astronomical events that occur over shorter timescales\cite{bellmGiveMeFew2022} or require additional follow-up observations. Here, FlyEye's ability to perform rapid, frequent scans of the sky complements the Vera-Rubin survey by providing immediate data on these transient phenomena.
Moreover, the Zwicky Transient Facility, which focuses on capturing events occurring over timescales from minutes to hours, also benefits from FlyEye's operations. FlyEye's design allows it to quickly adjust and point towards any part of the sky, providing crucial follow-up observations that help to confirm and better understand the initial findings from ZTF.
Through these collaborations, FlyEye not only fills crucial observational gaps but also enhances the collective scientific yield. It enables the capture and study of transient phenomena such as supernovae explosions, variable stars, and fast-moving near-earth objects, significantly contributing to our understanding of these dynamic celestial events. The combination of these facilities forms a robust network that can offer a comprehensive view of the universe's dynamic and transient nature, paving the way for breakthrough discoveries in astronomy.
\begin{table}[h]
\centering
\caption{Comparison of Telescope Specifications}
\label{tab:telescope_specs}
\begin{tabular}{|l|c|c|c|}
\hline
\textbf{Quantity} & \textbf{Current FLYEYE} & \textbf{ZTF} & \textbf{Vera Rubin} \\ \hline
FoV & $6.7 \times 6.7$ deg = 44 square deg & $7.2 \times 7.4$ deg = 47 square deg & 9.6 square deg \\ \hline
N pixels & $16 \times 4096 \times 4096 = 0.26$ billion & $16 \times 6144 \times 6160 = 0.6$ billion & 3.2 billion (total) \\ \hline
Gaps & No & Yes & Minimal \\ \hline
(E)ntendue & 0.00674 & 0.01359 & 0.0831 \\ \hline
Pixel Scale & 1.5 arcsec/px & 1.0 arcsec/px & 0.2 arcsec/px \\ \hline
\end{tabular}
\end{table}
\[
E = A \times \Omega,
\]
\[
A = \pi \left( \frac{D_{\text{effective}}}{2} \right)^2
\]
The entendue, $E$, is a function of the solid angle, $\Omega$, and of the effective collecting area, $A$, (0.5, 0.95, 35\,$m^2$ respectively).

The table above also shows that the optimal upgrade of the current FlyEye should include a larger effective diameter.

\section{Time-Domain Astronomy}
Traditional telescopes have provided snapshots that may miss critical changes or phenomena that occur rapidly or irregularly. However, FlyEye's rapid scanning capability addresses this challenge effectively.

Each night, FlyEye can scan two-thirds or the entire visible sky, capturing data across a wide spectrum of celestial events. This capability is invaluable for monitoring astronomical phenomena that exhibit variability over short periods, ranging from minutes to days. Examples include studying the development of supernovae from the initial explosion to the afterglow, capturing the sudden flare-ups of novae, or observing the interactions within binary star systems.

Specific design upgrades or observing strategies of the current FlyEye for Space Situational Awareness (SSA) extend beyond the primary mission goal of tracking near-Earth objects. The NEOSTEL design is optimized for the precise identification of asteroids on potential collision courses with Earth, as well as for studying their size, composition, and orbital patterns. Such capabilities are crucial for planetary defense initiatives and contribute significantly to the broader field of planetary science.

FlyEye's contributions to time-domain astronomy also extend to the detection of fast radio bursts and other high-energy phenomena. Its ability to quickly pivot and focus on different regions of the sky allows astronomers to follow up on alerts from other observatories, verifying and studying these puzzling bursts in real-time.

The design of the FlyEye telescopes not only complements existing observational frameworks but also significantly enhances the capabilities of global astronomy. By providing a dynamic and continuous view of the sky, it enables an unprecedented level of engagement with transient and variable celestial phenomena, opening new avenues for discovery and enriching our understanding of the universe’s complex and ever-changing nature.
\section{Enhanced Science Cases for Time-Domain Astronomy with FlyEye}

FlyEye's capabilities open up a wide array of possibilities for time-domain astronomy, enabling diverse scientific investigations that leverage its rapid scanning capability and wide field of view. One particularly exciting application is the follow-up observations of gravitational wave events. FlyEye can swiftly respond to alerts from gravitational wave detectors like LIGO\cite{collaborationAdvancedLIGO2015} and Virgo\cite{accadiaVirgoLaserInterferometer2012}, aiming to identify electromagnetic counterparts of these colossal cosmic collisions. This ability not only enriches our understanding of the physics of neutron star collisions and black hole mergers but also provides crucial insights into the extreme conditions influencing general relativity and quantum mechanics.

Moreover, the telescope is well-suited for exoplanet transit photometry, where it can monitor multiple stars simultaneously for dips in brightness caused by transiting exoplanets. This capability could dramatically increase the catalog of known exoplanets and offer a treasure trove of data for analyzing planetary atmospheres and compositions, particularly for those in the habitable zone.

FlyEye also promises advancements in the monitoring of Active Galactic Nuclei (AGN). By continuously observing these dynamic regions, astronomers can probe the physical processes at play around supermassive black holes, shedding light on galaxy evolution and black hole growth. This continuous monitoring extends to tracking from NEOs to the other solar system bodies.

Another area where FlyEye stands out is in the study of stellar phenomena such as flares. The telescope’s ability to capture sudden stellar brightening events across a broad expanse of the sky can provide valuable data on stellar magnetic activities and the mechanics of stellar atmospheres, enriching our understanding of stellar life cycles and evolution.

The panorama of FlyEye’s contributions to time-domain astronomy is vast, ranging from planetary science to the deep cosmos. To fully exploit this potential, collaborations with other observatories and the development of advanced data analysis techniques, including machine learning, are crucial. These efforts will ensure that FlyEye not only complements existing observational frameworks but also significantly enhances the capabilities of global astronomy, paving the road for discovery and enriching our understanding of the universe’s dynamic and ever-changing nature.

\section{CONCLUSIONS - 
Enhancing Our Understanding of the Dynamic Universe
}
A dedicated FlyEye for astronomy would have the opportunity to become a valuable instrument in observational astronomy, helping to unlock new frontiers in the study of the dynamic universe. 
By providing nightly coverage of the sky, FlyEye helps build a continuous and detailed record of the astronomical landscape. This capability is a strong asset not only for capturing rare astronomical events but also for tracking slower changes in the universe that are no less scientifically important. Such monitoring can lead to breakthroughs in understanding the structure and evolution of the universe.
The deployment of the FlyEye telescopes and further advancement as the new Mezzocielo design\cite{cerpelloniMezzocieloOneMeter2020,ragazzoniMezzocieloAttemptRedesign2020,ragazzoniCurrentStatusMezzoCielo2022,ragazzoniFlyEyeFamilyTree2023} thus represents a viable path for providing tools for the astronomers studying the temporal aspect of the cosmos. 
FlyEye telescope solution can extend the capabilities of foreseen astronomical surveys, thereby enriching our understanding of the dynamic and ever-changing universe.

\acknowledgments 
This work benefits of the discussions made within the ASI-INAF Contract N. 2023-50-HH.0 “Detriti spaziali e sostenibilità delle attività spaziali a lungo-termine”.

\bibliography{better} 
\bibliographystyle{spiebib} 

\end{document}